# Plasmonic enhancement of the third order nonlinear optical phenomena: figures of merit

Jacob B. Khurgin[1,*] and Greg Sun[2]

[1]*Department of Electrical & Computer Engineering, Johns Hopkins University, Baltimore, MD 21218, USA*
[2]*Department of Physics, University of Massachusetts at Boston, Boston, MA 02125, USA*
[*]*jakek@jhu.edu*

**Abstract:** Recent years have seen increased interest to plasmonic enhancement of nonlinear optical effects, yet there remains an uncertainty of what are the limits of this enhancement. We present a simple and physically transparent theory of plasmonic enhancement of third order nonlinear optical processes achieved in plasmonic structures and show that while huge enhancement of effective nonlinear index can be attained, the most relevant figure of merit, the of phase shift per one absorption length remains very low. This means that while on one hand nonlinear plasmonic materials are not well suitable for applications requiring high efficiency, e.g. all-optical switching and wavelength conversion, on the other hand they can be very useful for the applications where the overall high efficiency is not a must, such as sensing.

## References and links

# 1. Introduction

Nonlinear optical phenomena have been in the focus of interest of the scientific community ever since the scientists gained access to intense optical fields, i.e. since the invention of laser in 1960 [1]. Indeed, shortly after this invention all the major nonlinear optical phenomena of second and third order have been demonstrated [2-4] and the theory of nonlinear optics has been developed [5-7]. Today a clear understanding of the nonlinear optical effects in various media exists [8, 9]. The fascinating promise of nonlinear optics has always been based on the fact that nonlinear optical phenomena allow one in principle to manipulate photons with other photons without relying on electronics. And yet, while there have been some spectacular success stories that lead to practical products (such as frequency converters, Optical Parametric Oscillators, frequency, and a few others), the majority of nonlinear optical phenomena so far have not become competitive for practical applications, simply because the magnitude of fast nonlinear effects is small.

One may recall that all nonlinear optical phenomena can be divided into two broad classes: slow and ultra-fast. The slow nonlinear phenomena are generally classified as such by the fact that optical fields do not interact directly, but through the various "intermediaries", such as electrons excited when the photons get absorbed, or through the temperature rise caused by the release of the energy of the absorbed photons. For as long as these "intermediaries" exist, i.e. while the electrons stay in the excited state or until the heat dissipates, their effect on the optical fields accumulates, hence these phenomena, such as saturable absorption, photo-refractive effect, or thermal nonlinearity, can be quite strong, but this very fact makes them slow, as their temporal response is limited by a time constant associated with appropriate relaxation, recombination, or heat diffusion times.

The other, so called ultrafast nonlinearities, do not involve excitation of the matter to the real excited states as there exist no transitions between the states that are resonant with the photon energy, hence they often referred to as "virtual". When the non-energy-conserving "virtual" excitation does take place its duration is determined by the uncertainty principle, and thus can be as short as a few femtoseconds or even a fraction of femtosecond which explicates the term "ultra-fast". But it is precisely the fact that the excitation lasts such a short time interval that makes the ultra-fast nonlinearities relatively weak. For example, the nonlinear refractive index, $n_2$ that characterizes third order nonlinearities, ranges from $n_2 \sim 5 \times 10^{-16} \text{cm}^2/\text{W}$ for fused silica that is transparent all the way to UV, to perhaps $n_2 \sim 1 \times 10^{-13} \text{cm}^2/\text{W}$ for chalcogenide glasses transparent only in the IR range [10].

Therefore, very strong optical power density on the order of GW/cm$^2$ is required in order to produce appreciable ultrafast nonlinear optical phenomena. The average optical power available from a compact laser rarely exceeds a few hundred milliwatts, furthermore, if one wants to envision all optical integrated circuits, the power dissipation requirements constrain the power to even much lower levels than that, possibly less than a milliwatt. Hence early on it was understood that to make nonlinear optical phenomena practical one must concentrate the power in both space and time. Concentration in space usually implies coupling the light into tightly-confining optical waveguide or a fiber. But the attainable concentration is limited to roughly a wavelength in the medium by the diffraction limit. In addition, one may consider resonant concentration of optical energy in micro-cavities [11-12], ring resonators [13], photonic bandgap structures [14] and slow light devices [15,16], but all the resonant effect inevitably limit the bandwidth [17]. It is the concentration of optical power in time domain provided by pulsed sources, particularly by the Q-switched [18] and mode-locked lasers [19,20], that has proven to be the winning technique in nonlinear optics. In low duty cycle mode-locked pulse the peak power exceeds the the average power by may orders of magnitude hence use of ultra-short low duty cycle pulses has become ubiquitous method of obtaining excellent practical results both for the second and especially third order (optical frequency comb and continuum generation) phenomena. And yet if one is thinking of applications in information processing, the switches are expected to operate at the same symbol rate and duty cycle as the data stream. Then one should look at other methods of concentrating the energy and one's attention is inevitably drawn back to the space domain and the question arises: can one transfer the mode-locking techniques from time to space, i.e. to create a low duty cycle high peak power distribution of optical energy in space, rather than in time and to use it to effectively enhance nonlinear optical effects.

Extending the time space analogy, let us look at what limits the degree of energy concentration in time and space. In time domain it is obviously dispersion of group velocity, while in space domain it is the diffraction. While there is obvious equivalence between the mathematical description of dispersion and diffraction, there is stark difference – the group velocity dispersion can be minimized by a number of techniques because it can be either positive or negative, while the diffraction is always positive and there exists a hard diffraction limit to optical confinement in all dielectric medium. But, of course, the diffraction limit is applicable only to the all-dielectric structures with positive real parts of dielectric constant. In all-dielectric structures the energy oscillates between electric and magnetic fields, and if the volume in which one tries to confine the optical energy is much less than a wavelength the magnetic field essentially vanishes (so-called quasi-static limit) and without this energy "reservoir" for storage every alternative quarter-cycle the energy simply radiates away. But if the structure contains medium with negative dielectric constant (real part), i.e. free electrons, an alternative reservoir for energy opens up – the kinetic motion of these free carriers in metal or semiconductor, and the diffraction limit ceases being applicable. The optical energy can be then contained in the tightly confined sub-wavelength modes surrounding or filling the gap between the tiny metallic particles. These modes, combining electric field with charge oscillations are called localized surface plasmons (LSP) and in the last decade the whole new fields of plasmonics and closely related metamaterials arise with the ultimate goal of taking advantage of the unprecedented degree of optical energy concentration on the sub-wavelength scale [21].

In the last decade researchers have observed enhancement of both linear (absorption, luminescence) [22,23] and non-linear (Raman) phenomena [24-26] in the vicinity of small metal nanoparticles and their combinations. Experimentally, surface-enhanced Raman scattering, enhancement of many orders of magnitude has been observed [24-26], while the enhancement for luminescence and absorption was more modest. To address this issue we have developed a rigorous yet physically transparent theory explaining this enhancement provided by single [27] or coupled [28,29] nanoparticles in which we have traced relatively weak enhancement of luminescence to large absorption in the metal, which cannot be reduced in truly subwavelength mode in which the field is concentrated [30,31]. In that work [30] we have shown that the decay rate of the electric field in the sub-wavelength mode is always on the order of the scattering time in metal, i.e. 10-20 fs in noble metals. This is the natural consequence of the aforementioned fact that half of the time all the energy is contained in the kinetic motion of carriers in the metal where it dissipates with the scattering rate. As a result, a significant fraction of the SP's simply dissipates inside the metal rather than radiating away. The net result is that only very inefficient emitters [32] and also absorbers [33] can be enhanced by plasmonic effects, such as, of course, the Raman process that is extremely inefficient [34], while the relatively efficient devices, such as LED [35], solar cells [36], and detectors [37] do not exhibit any significant plasmonic enhancement relative to what can be obtained without the metal by purely dielectric means [38].

Therefore, it is only natural to investigate what plasmonic enhancement can do for the inherently weak nonlinear processes, and, although the first works along this direction are over 30 years [39-44] the interest has peaked up significantly in the last decade [45]. There are a number of ways where nonlinear optical effects can be enhanced by the surface plasmons. One is the coupling of excitation field to form the much stronger localized field near the surface of metal structure that leads to the enhancement of optical processes [46]. Such a strong near-field effect is responsible for the experimental observations of significant Raman enhancement that has resulted in single molecule detection [24-26,47], surface plasmon enhanced wave mixing like SHG on random [48-50] and defined plasmonic structures [51-57], as well as the enhancement of linear processes such as optical absorption and luminescence [22,23]. Another is fact that surface plasmon resonance is ultra-sensitive to the dielectric properties of the metal and its surrounding medium – a minor modification in the refractive index around the metal surface can lead to a large shift of plasmonic resonance [58]. Such a phenomenon brings about the prospect of controlling light with another light where the latter induces optical property changes in the plasmonic structure which in turn modifies the propagation of the original light. Motivated by this promise, researchers around the world have been pursuing the goal of practical all optical modulation or switching based on Kerr nonlinearities in either unconfined plasmonic materials [59-62] or waveguides [63-67], which has remained elusive up to this date.

At this point it is important to differentiate between the sources of nonlinearity in these works, because both metals and dielectrics possess nonlinearity. The nonlinear susceptibility of the metals can be due to either free carriers or due to band-to-band transitions. The nonlinearity of band-to-band transition (typically involving d-bands in noble metals) is no different from the interband nonlinearity of dielectrics and semiconductors, except it always occurs in the region of large absorption due to free carriers, and, on top of it, the nonlinearity is strongest in the blue region of spectrum, while we prefer to concentrate on the telecommunication region of 1300-1500nm. As far as nonlinearity of free electrons goes, it is extremely weak because LSP's (at least when there are only a few of them per nanoparticle) are nearly perfect harmonic oscillators. Therefore, we shall consider the structure in which the metal nanoparticles are embedded into the nonlinear material with large nonlinearity and low loss. We shall limit our consideration to the third order nonlinearity because it leads to optical switching and other interesting phenomena without phase-matching, and, furthermore, we shall limit ourselves to the nonlinear modulation of the refractive index (real part of susceptibility) rather than absorption (imaginary part). One reason for it is that for the amplitude modulation it is desirable to maintain the "zero" bit level as close to real zero as possible, which can only be done by the interference (as in, for instance, Mach Zehnder interferometer). Another reason is that by modulating index one can take advantage of advanced phase –modulation formats, such as quadrature phase-shift keying (QPSK), quadrature amplitude modulation (QAM) etc. Index modulation is typically broadband and, in addition to simple modulation and switching, can be used for frequency conversion, while absorption modulation is an inherently resonant phenomenon. Finally, changes in absorption are usually associated with real excitations hence they are not truly ultrafast.

We consider the structures shown in Fig. 1(a,b) in which nonlinear dielectric surrounds the metal nanoparticles. The goal of our treatment is to evaluate the enhancement of the third order nonlinear polarizability of this metamaterial, or one can use the term "artificial dielectric" consisting of metal nanoparticles that enhance local field. In the course of this work we shall introduce relevant figures of merit relevant to practical applications and see how the plasmonically enhanced nonlinear materials stack up against the conventional ones. To make our treatment both general and physically transparent we shall fully rely on analytical derivations, which, of course, would require certain simplifications, that are justified for as long as one is looking just for the order of magnitude of enhancement. For instance, we shall consider just spherical or elliptical (or spheroidal) nanoparticles, single and coupled, but we shall indicate how the treatment can be expanded to other shapes of nanoparticles, including nanoshells [68], that can be defined by just three parameters: resonant SP frequency $\omega_0$, quality factor $Q$, and effective SP mode volume $V_{eff}$. For this purpose, In Fig. 1(a) we show spherical nanoparticles and in Fig. 1(b) we show the elliptical nanoparticle with resonance at telecommunication wavelength of 1320 nm with actual field distribution calculated numerically. Also shown in Fig.1(c) the extinction spectrum of the elliptical particle where the resonance can be observed.

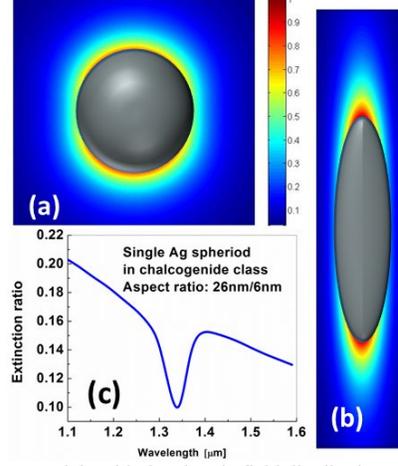

Fig. 1. (a) Spherical Ag nanoparticle with the electric field distribution. (b) Elliptical nanoparticle resonant at 1320 nm and associate electric field distribution (c) Extinction spectrum of the above nanoparticle

## 2. Isolated metal nanoparticles embedded in the dielectric: linear properties

Consider a rather general scheme for plasmoncially enhanced nonlinearity shown in Fig. 2(a) consisting of nanospheres of radius $a$ surrounded by the nonlinear dielectric with relative permittivity $\varepsilon_d$ and nonlinear susceptibility tensor $\chi^{(3)}$. The concentration of spheres is $N_s$. In most general case $\chi^{(3)}$ implies four wave interactions, with some of the waves being the pumps (of switching signals) and some being the nonlinear output signals. In many practical cases, such as cross- and self-phase modulation there is degeneracy and the number of interacting waves is reduced. In Fig. 2(a) we show just one pump (or switching) wave of frequency $\omega$ and one signal wave of frequency $\omega'$.

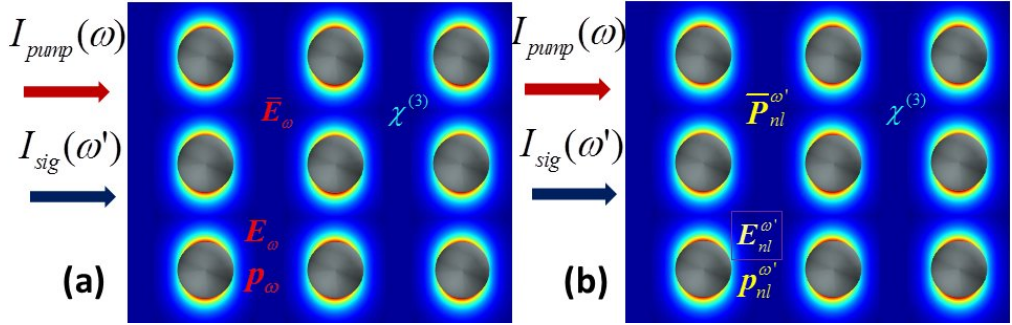

Fig. 2. Fields and polarizations in the plasmonically enhanced nonlinear metamaterial (a) Average $\bar{E}_\omega$ and local $E_\omega$ electric fields and dipole $p_\omega$ at the pump frequency. (b) local nonlinear field $E_{\omega'}$, dipole moment $p_{nl}^{\omega'}$ and average nonlinear polarization $\bar{P}_{nl}^{\omega'}$

As the pump wave propagates through the material, the average electric field is $\bar{E}_\omega$ and in this field the nanospheres become polarized, i.e. acquire the dipole moment [69]

$$\boldsymbol{p}_\omega = \frac{\varepsilon_m - \varepsilon_d}{\varepsilon_m + 2\varepsilon_d} 4\pi\varepsilon_0\varepsilon_d a^3 \overline{\boldsymbol{E}}_\omega \qquad (1)$$

as shown in Fig. 2(a). Using Drude model for the dielectric constant of metal $\varepsilon_m = 1 - \omega_p^2 / (\omega^2 + j\omega\gamma)$ with plasma frequency $\omega_p$ and scattering rate $\gamma$ we can obtain

$$\boldsymbol{p}_\omega = \frac{\alpha\omega_0^2 \overline{\boldsymbol{E}}_\omega}{\omega_0^2 - \omega^2 - j\omega\gamma} \approx \alpha \frac{Q}{L(\omega)} \overline{\boldsymbol{E}}_\omega \qquad (2)$$

where $\omega_0 = \omega_p / \sqrt{1+2\varepsilon_d}$ is the LSP resonant frequency [32], $Q = \omega_0/\gamma$ is the $Q$-factor of the mode $L(\omega) = Q(1 - \omega^2/\omega_0^2) - j$ is resonant Lorentzian denominator, $\alpha \approx 3\varepsilon_0\varepsilon_d V\beta$ is the polarizability of the nanoparticle and $\beta = 3\varepsilon_d / (2\varepsilon_d + 1)$. For particles of different shapes $\beta$ will be somewhat different and polarization-dependent, yet still within the same order of magnitude. Similarly, the value of resonant frequency will change, however, since we are interested only in the order of magnitude results in this work, all the conclusions obtained here for spherical particles and their combinations will hold for the particles of different shapes. It should be noted that the $Q$-factor for different shape is depends only on the resonant frequency $\omega_0$ since the decay rate $\gamma$ does not depend on the shape (or exact dimensions) as long as particles are much smaller than wavelength (which is of course required to avoid scattering and diffraction effects).

The Q factor for the gold and silver, two lowest loss plasmonic materials are shown in Fig. 3 as functions of frequency. Near 1320 nm Q-factor of bulk gold is about 12 and for the bulk silver it is closer to 30 according to Johnson and Christy [70], although for the silver nanoparticles the interface scattering usually decreases the Q factor by a factor of few. Also, gold is easier to work with than silver, as it does not get oxidized, so majority of researchers use gold in the telecom region. In this work, however, we consider the best case scenario and hence use silver as an example with Q=20 which is probably on the higher side of most experimental results but our goal is to look at the best case scenario.

Equation (2) can be construed as the solution of the equation of motion of the harmonic oscillator, or the LSP mode characterized by the dipole moment $\boldsymbol{p}$

$$\frac{d^2\boldsymbol{p}}{dt^2} + \gamma\frac{d\boldsymbol{p}}{dt} = -\omega_0^2\boldsymbol{p} + \omega_0^2\alpha\overline{\boldsymbol{E}} \qquad (3)$$

and consisting of coupled oscillations of the free electron current insider inside the nanoparticle, and the electric field [32]

$$\boldsymbol{E}_\omega(\boldsymbol{r}) = \begin{cases} -\dfrac{\boldsymbol{p}}{4\pi\varepsilon_0\varepsilon_d a^3} & r < a \\ \dfrac{1}{4\pi\varepsilon_0\varepsilon_d r^3}\left[3(\boldsymbol{p}\cdot\hat{\boldsymbol{r}})\hat{\boldsymbol{r}} - \boldsymbol{p}\right] & r > a \end{cases} \qquad (4)$$

inside and outside the nanoparticle, respectively, with the maximum field near the surface of nanoparticle equal to

$$\boldsymbol{E}_{\max,\omega} \approx \frac{2\beta Q}{L(\omega)} \overline{\boldsymbol{E}}_\omega. \qquad (5)$$

Hence near the resonance the local field is enhanced roughly by a factor $2Q$ relative to the average field.

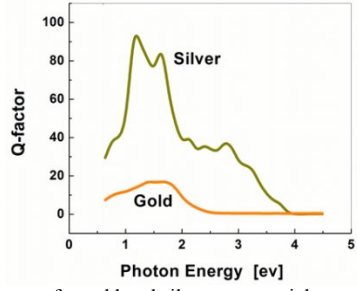

Fig. 3. Dispersions of Q-factors for gold and silver nanoparticles.

If the nanoparticles are much smaller than the wavelength of light in the dielectric, one can apply a classical polarizability theory in which each nanoparticle is treated as polarizable atom. The effective dielectric constant of the composite medium (or a metamaterial if one wants to use a more modern, de rigueur terminology) can be found as the sum of the original dielectric constant and the susceptibility of the nanoparticles with a density $N_s$,

$$\varepsilon_{eff} = \varepsilon_d + \frac{N_s \alpha}{\varepsilon_0} \frac{Q}{L(\omega)} = \varepsilon_d \left(1 + 3f\beta \frac{Q}{L(\omega)}\right) \qquad (6)$$

where we have introduced the effective filling factor $f = N_s V \ll Q^{-1}$. The latter condition is practically always satisfied in the medium with Q~20 and is required to avoid taking into account dipole-dipole interaction effects that would change the LSP resonant frequency according to Lorentz-Lorentz formula. But, once again, even for very dense medium frequency renormalization is not going to change the main conclusions of this work.

In this approximation we can find the effective index of refraction

$$n_{eff} = \varepsilon_{eff}^{1/2} \approx n_d \left[1 + \frac{3f\beta}{2} \frac{Q^2(1 - \omega^2/\omega_0^2)}{|L(\omega)|^2} + \frac{3f\beta}{2} \frac{jQ}{|L(\omega)|^2}\right], \qquad (7)$$

where $n_d = \sqrt{\varepsilon_d}$. Obviously, the effective absorption coefficient is

$$\alpha_a = \frac{2\pi n_d}{\lambda} \frac{3f\beta Q}{|L(\omega)|^2}, \qquad (8)$$

And it also gets resonantly enhanced by the *Q*-factor.

## 3. Isolated metal nanoparticles embedded in the dielectric: third order non-linear properties

### 3.a Nonlinear polarization and effective susceptibility

Let us now turn our attention to Fig. 2b where the local nonlinear microscopic polarization at the frequency $\omega'$,

$$\boldsymbol{P}_{nl}(\boldsymbol{r}, t) = P_{max,nl}^{\omega'} \boldsymbol{G}(\boldsymbol{r}) e^{-j\omega' t} \qquad (9)$$

is established near the nanoparticle due to the presence of strong local pump field. As mentioned above, $\omega'$ could be the same as or different from the pump frequency $\omega$ that drives the nonlinear polarization. The maximum nonlinear polarization, usually occurring at the same location where the local pump field reaches maximum $|\boldsymbol{P}_{nl}(\boldsymbol{r}_{max})| = P_{max,nl}^{\omega'}$ and $\boldsymbol{G}(\boldsymbol{r})$ is the normalized shape of nonlinear polarization. The nonlinear polarization can now drive the LSP oscillations at the same frequency $\omega'$ according to the wave equation for the electric field of the LSP mode

$$\nabla^2 \boldsymbol{E}(\boldsymbol{r},t) - \frac{\varepsilon_r(\boldsymbol{r})}{c^2}\frac{\partial^2}{\partial t^2}\boldsymbol{E}(\boldsymbol{r},t) = \frac{1}{\varepsilon_0 c^2}\frac{\partial^2}{\partial t^2}\boldsymbol{P}_{nl}(\boldsymbol{r},t) \;. \tag{10}$$

We look for the solution of the form

$$\boldsymbol{E}(\boldsymbol{r},t) = \sum_l E_{max,l}^{\omega'} \boldsymbol{F}_l(\boldsymbol{r}) e^{-j\omega't} \tag{11}$$

where $\boldsymbol{F}_l(\boldsymbol{r})$ is the normalized electric field of the $l$-th LSP eigen-mode with $l=1$ being the dipole mode described by (4), whose amplitude $E_{max,1}$ we are trying to determine. Substituting (11) into (10) and using modes orthogonality we obtain for the steady state amplitude of the $l=1$ dipole mode driven by the nonlinear polarization at frequency $\omega'$

$$E_{max}^{\omega'} = \frac{P_{max,nl}^{\omega} \kappa}{\varepsilon_0 \varepsilon_d} \frac{Q}{L(\omega')} \;. \tag{12}$$

where the overlap coefficient, assuming that dielectric is non-dispersive and non-lossy, is

$$\kappa = \varepsilon_d \int \boldsymbol{F}_1(\boldsymbol{r}) \cdot \boldsymbol{G}(\boldsymbol{r}) dV \;/\; \int \frac{\partial(\varepsilon_r'\omega)}{\partial \omega} F_1^2(\boldsymbol{r}) dV \;. \tag{13}$$

Now, according to (2) we can find the amplitude of the dipole mode as

$$p_{nl}^{\omega'} = \frac{3}{2} V \varepsilon_0 \varepsilon_d E_{max,nl}^{\omega'} = \frac{3}{2} V \kappa \frac{Q}{L(\omega)} P_{max,nl}^{\omega'} \;. \tag{14}$$

and overall **effective nonlinear polarization of the metamaterial** is

$$\overline{\boldsymbol{P}}_{nl}^{\omega'} = \frac{3}{2} f \kappa \frac{Q}{L(\omega)} \boldsymbol{P}_{max,nl}^{\omega'} \tag{15}$$

as shown in Fig. 2(b). As one can see, local nonlinear polarization gets enhanced by being at resonance with the nanoparticle dipole mode and enhancement is once again proportional to the $Q$-factor of the resonance.

It is instructive to re-cap the chain of events that leads to establishment of enhanced nonlinear polarization as shown in Fig. 2:

i. The average incoming pumping field $\overline{\boldsymbol{E}}_\omega$ polarizes nanoparticles engendering linear dipole moment $\boldsymbol{p}_\omega$ in each of them;

ii. Dipole oscillations are coupled with linear local field $\boldsymbol{E}_\omega(\boldsymbol{r})$ in the vicinity of each nanoparticle. This field is resonantly enhanced by a factor of the order of $Q$ relative to $\overline{\boldsymbol{E}}_\omega$;

iii. A local nonlinear polarization $\boldsymbol{P}_{nl}^{\omega'}(\boldsymbol{r})$ is established in the vicinity of each nanoparticle. Since this polarization is proportional to the third order of field, it is enhanced roughly by a factor of $Q^3$;

iv. This polarization resonantly couples into the dipole LSP mode of the nanoparticle thus establishing the local nonlinear field $\boldsymbol{E}_{\omega'}(\boldsymbol{r})$ and dipole moment $\boldsymbol{p}_{nl}^{\omega'}$. Resonance causes enhancement by another $Q$-factor;

v. Finally the localized dipoles $p_{nl}^{\omega'}$ combine to establish the average nonlinear polarization $\overline{P}_{nl}^{\omega'}$, enhanced by $\sim Q^4$ relative to the nonlinear polarization in the absence of nanoparticles.

Needles sto say, all the steps outlined above occur simultaneously and instantly, but in our view tracing the process step by step is instructive as it reveals the physical picture.

We now turn our attention to specifically third order processes. Consider the third order nonlinearity in which interaction of electromagnetic waves at three different frequencies described by the general local third order susceptibility.

$$P_{nl}^{\omega_1-\omega_2+\omega_3}(r) = \varepsilon_0 \chi^{(3)}(\omega_3, -\omega_2, \omega_1) E_{\omega_1}(r) E_{\omega_2}^*(r) E_{\omega_3}(r) . \tag{16}$$

In general, when all four frequencies, $\omega_1$, $\omega_2$, $\omega_3$, and $\omega_4 = \omega_1 - \omega_2 + \omega_3$ are different (but typically close to each other) the nonlinear process described by (16) is four wave mixing (FWM), when $\omega_3 = \omega_1$ $\omega_4 = 2\omega_1 - \omega_2$ (16) describes optical parametric generation (OPG), when $\omega_1 = \omega_2$ and $\omega_3 = \omega_4$ it describes cross-phase modulation (XPM) and for the case when all frequencies are equal (16) describes self-phase modulation (SPM). FWM and OPG are both of great interest in wavelength conversion while both XPM and SPM are important for optical switching.

In a composite medium the local fields $E_{\omega_k}(r)$ in (16) in the vicinity of the nanoparticle are all locally enhanced relative to the mean fields $\overline{E}_{\omega_k}$ according to (5), i.e.

$$E_{\omega_k}(r) = \frac{2\beta Q}{L(\omega_k)} \overline{E}_{\omega_k} F_1(r) . \tag{17}$$

Hence the local third-order nonlinear polarization is

$$P_{nl}^{\omega_1-\omega_2+\omega_3}(r) = P_{\max,nl}^{\omega_1-\omega_2+\omega_3} G(r) \tag{18}$$

where the amplitude is

$$P_{\max,nl}^{\omega_1-\omega_2+\omega_3} = \varepsilon_0 \left|\chi^{(3)}(\omega_3, -\omega_2, \omega_1)\right| \frac{(2\beta Q)^3}{L(\omega_1) L^*(\omega_2) L(\omega_3)} \overline{E}_{\omega_1} \overline{E}_{\omega_2}^* \overline{E}_{\omega_3} , \tag{19}$$

while the shape function is

$$G(r) = \chi^{(3)} / \left|\chi^{(3)}\right| F_1(r) F_1(r) F_1(r) , \tag{20}$$

and $\chi^{(3)} / \left|\chi^{(3)}\right|$ is the normalized fourth-order nonlinear susceptibility tensor. Substituting (20) into (15) we obtain

$$\overline{P}_{nl}^{\omega_4} = \varepsilon_0 \chi_{eff}^{(3)}(\omega_3, -\omega_2, \omega_1) \overline{E}_{\omega_1} \overline{E}_{\omega_2}^* \overline{E}_{\omega_3} \tag{21}$$

where, for the typical case of all frequencies being close to each other, the effective nonlinear susceptibility is

$$\chi_{eff}^{(3)} \square \frac{3}{2} f \kappa_3 (2\beta)^3 \frac{Q^4}{L^2(\omega) \left|L(\omega)\right|^2} \chi^{(3)} \tag{22}$$

and the coupling coefficient for the third-order nonlinearity

$$\kappa_3 = \frac{\varepsilon_d}{V_{eff,1}} \int_{r>a} F_1(r) \cdot \frac{\chi^{(3)}}{\left|\chi^{(3)}\right|} F_1(r) F_1(r) F_1(r) d^3r . \tag{23}$$

The nonlinear susceptibility thus gets enhanced by a factor proportional to $Q^4$. This is an encouraging result, exciting enough to draw attention of both plasmonic and nonlinear optics communities to this topic, which has witnessed an upsurge of research efforts and publications as described in Introduction. Indeed, even with $f \sim 0.001$ filling ratio one can expect more than a 100-fold enhancement of susceptibility and nonlinear refractive index and indicates that one can achieve the same efficiency of nonlinear phase modulation at less 1/100 of the length of conventional device, and, more dramatically, same efficiency of the wavelength conversion in less than 1/10,000 of the length! It is these results that are often quoted as justification for using nanoplasmonics to enhance nonlinearity, yet one needs to maintain caution when it comes to reporting these giant plasmonic enhancements. Our prior research of plasmonic enhancement of various emission processes including photoluminescence [27], electroluminescence [32] and Raman scattering [34] has shown that large enhancements are feasible only for the processes that have very low original efficiency (such as Raman scattering) but are far more modest for the efficient processes such as fluorescence and electroluminescence. It is therefore reasonable to expect that there must exist an upper limit of the nonlinear plasmonic enhancement.

### 3.b Effective nonlinear index and maximum phase shift

To understand the limitations of the enhancement we shall first consider XPM (or SPM) case for which nonlinear polarization in (16) can be written as $P_{nl}^{\omega_2}(\boldsymbol{r}) = 2\varepsilon_0 n_d n_2 I_{\omega_1}(\boldsymbol{r}) E_{\omega_2}(\boldsymbol{r})$ where $n_2 = \chi^{(3)} \eta_0 / \varepsilon_d$ is the nonlinear index of the dielectric, and $I_{\omega_1}(\boldsymbol{r}) = |E_{\omega_1}(\boldsymbol{r})|^2 n_d / 2\eta_0$ is the local intensity  Similarly, we now introduce the effective nonlinear index as $n_{2,\text{eff}} = \chi_{\text{eff}}^{(3)} \eta_0 / \varepsilon_d$ and write average nonlinear polarization as $\overline{P}_{nl}^{\omega_2} = 2\varepsilon_0 n_d n_{2,\text{eff}} \overline{I}_{\omega_1} \overline{E}_{\omega_2}$ According to (22) the effective nonlinear index gets enhanced by the same giant factor proportional to $Q^4$,

$$n_{2,\text{eff}} \sim \frac{3}{2} f \kappa_3 (2\beta)^3 \frac{Q^4}{L^2(\omega_2) |L(\omega_1)|^2} n_2. \tag{24}$$

Next we estimate the nonlinear phase shift in the absorbing medium as

$$\Delta\Phi(z) = \frac{2\pi}{\lambda} n_{2,\text{eff}} \int_0^z \overline{I}_{\omega_1}(z) dz = \frac{2\pi}{\lambda \alpha_a} n_{2,\text{eff}} \overline{I}_0 (1 - e^{-\alpha_a z}) \tag{25}$$

where $\overline{I}_0$ is the input pump intensity, $\alpha_a$ is the absorption coefficient defined in (8). This means that the maximum phase shift obtained after about one absorption length is

$$\Delta\Phi_{\max} = \frac{|L(\omega_1)|^2}{3 f \beta Q} \frac{n_{2,\text{eff}}}{n_d} \overline{I}_0 \approx \kappa_3 (2\beta)^2 \frac{Q^3}{L^2(\omega_2)} \frac{n_2}{n_d} \overline{I}_0. \tag{26}$$

Achieving the π-phase shift required to get photonic switching would then require at resonance $\overline{I}_\pi \sim \pi n_d \left[ \kappa_3 n_2 (2\beta)^2 Q^3 \right]^{-1}$. If we assume $\beta \sim 1.45$ (estimated numerically for the actual ellipsoid resonant at 1320 nm of Fig. 1(b), $Q \sim 20$ and large nonlinear index characteristic of chalcogenide glass $n_2 = 10^{-13} \text{cm}^2 / W$, the required switching intensity is then on the order of $\overline{I}_\pi \sim 1.6 \times 10^9 \, W/cm^2$, which is quite high.   The implication is that the giant nonlinear index enhancement (24) can only be used to reduce the length of the device, while the switching intensity remains quite high – requiring peak powers of about tens of W into 1μm² waveguide.

But next we shall ask the question – what is the actual local intensity near the metal surface? According to (5) the maximum local intensity approaches $I_{max} = (2\beta Q)^2 \overline{I}_\pi \approx 5 \times 10^{12}$ W/cm$^2$ (local field in excess of $5 \times 10^7$V/cm, which is significantly higher than damage threshold of the material. In fact, if one searches through all the nonlinear materials, it is difficult to find one that is capable of achieving ultrafast refractive index change larger than 0.1%. In addition to limitation due overheating and optical damage, at high power the nonlinearities of higher than third order, i.e. $\chi^{(5)}$, $\chi^{(7)}$ become important, and they often have their sign opposites to $\chi^{(3)}$ [71] which leads to actual decrease in the nonlinear index change at high intensities [72].

Therefore, let us define the maximum local nonlinear index change attainable in a given material as $\Delta n_{max}$ and obtain the maximum change of effective index

$$\Delta n_{eff,max} = n_{2,eff} \overline{I} \approx 3 f \kappa_3 \beta \frac{Q^2}{L^2(\omega_2)} \Delta n_{max} . \tag{27}$$

As we can see now the enhancement is only proportional to $Q^2$. This result makes perfect sense if we recognize that local change of dielectric constant $\Delta \varepsilon_{d,max} = 2n_d \Delta n_{max}$ simply causes the shift of the LSP resonant frequency $\omega_0 = \omega_p / \sqrt{1 + 2\varepsilon_d}$, which in turn changes the effective dielectric constant of the metamaterial $\varepsilon_{eff,max}$ according to (6) proportionally to $Q^2$ as we differentiate the Lorentzian in (6). It is crucial to note that this factor of $Q^2$ in (27) is applicable not just to an isolated nanoparticle but also to more sophisticate structures, like dimers and nanoantennae – in each case the local change of index causes the shift of plasmonic resonance proportional to the same factor of $Q^2$

It follows then **that maximum obtainable phase shift** (26) can be found as

$$\Delta \Phi_{max} = \frac{2\pi}{\lambda \alpha_a} \Delta n_{eff,max} = \kappa_3 Q \frac{|L(\omega_1)|^2}{L^2(\omega_2)} \frac{\Delta n_{max}}{n_d} . \tag{28}$$

The simple meaning of (28) is that, even if we assume enormous local nonlinear index change of 1% (i.e. local intensity of $10^{11}$ W/cm$^2$ ), we cannot expect to get phase shift higher than 0.1, almost two orders of magnitude less than required for π-phase shift switching. Notice also that for closely spaced frequencies of pump and signal the maximum phase shift does not even depend strongly on the position relative to SPP resonance as increase in nonlinearity is balanced by the increase in absorption. It should be also noted that the expression (28) can be used independent of the origin of the index change, i.e. it does not have to be all optical but can also be electro-optical or thermo-optical.

### 3.c Efficiency of Frequency conversion

It is easy to see that small maximum phase shift for XPM or SPM corresponds to even smaller efficiency of the frequency conversion for FWM or OPG. Indeed the growth of the idler $\overline{E}_{\omega_3}(z)$ in the presence of pump $\overline{I}_{\omega_1}(z) = \overline{I}_0 e^{-\alpha_a z}$ and signal $E^*_{\omega_2}(z) = E_s e^{-\frac{\alpha_a}{2}z}$ can be found as

$$\overline{E}_i(z) = \frac{2\pi}{\lambda \alpha_a} n_{2,eff} \overline{I}_0 E_s \left[1 - e^{-\alpha_a z}\right] e^{-\alpha_a z/2} \tag{29}$$

with a maximum near $z = \alpha_a^{-1} \ln 3$ equal to

$$\overline{E}_{i,\max} = \frac{2\pi}{\lambda \alpha_a} \frac{2}{3^{3/2}} n_{2,eff} \overline{I}_0 E_s = \frac{2}{3^{3/2}} \Delta\Phi_{\max} E_s. \tag{30}$$

Therefore, maximum conversion efficiency from the signal to the idler is

$$I_i / I_s \sim 0.15 \Delta\Phi_{\max}^2 \tag{31}$$

and under no conceivable conditions it can exceed -30dB.

Here we should also briefly mention that one could use modulation of the refractive index of the metal itself, but it is difficult to see how one can change the index of metal by more than 1% unless one operates near the interband transitions where the $Q$-factor is greatly reduced which defeats the whole purpose of plasmonic enhancement.

### 4. Enhancement of nonlinearity in more complex structures: dimers or nanolens

#### 4.a Field Enhancement

Now we have concluded that while nonlinear susceptibility and nonlinear index of refraction do get enhanced significantly in the simple nanostructures, the strong absorption makes maximum attainable phase shift less than desired. From the previous work of our own [73,74] as well as from others [75], we have established that local fields can be enhanced even further in more complicated nanoparticle structures. We have shown that local fields in the gap between two identical nanoparticles (dimer) [73] or in the vicinity of a smaller nanoparticle coupled to a larger nanoparticle of the same shape (nanolens) [74]. We have shown that in both cases the maximum filed enhancement was proportional to $Q^2$ rather than $Q$ for a single nanoparticle, hence much larger "cascaded" enhancements of absorption, Raman scattering, and in some cases photoluminescence could be achieved in these "hot spots". Therefore, it is tempting to evaluate the possibility of using the hot spots to enhance nonlinearity. Since we have shown that in either dimer or nanolens the field enhancement is similar, we shall limit our analysis to the case of nanolens only, as it is easier to describe analytically.

Consider two spherical nanoparticles of radii $a_1$ and $a_2$ separated by a vector $\mathbf{r}_{12}$ as shown in Fig. 4(a). The dipole oscillation equation (3) is augmented by the dipole-dipole interaction between the two dipoles associated with the two coupled nanoparticles,

$$\frac{d^2 \mathbf{p}_{1(2)}}{dt^2} + \gamma \frac{d\mathbf{p}_{1(2)}}{dt} = -\omega_0^2 \mathbf{p}_{1(2)} + \omega_0^2 \alpha_{1(2)} \overline{\mathbf{E}}_\omega + \omega_0^2 \alpha_{1(2)} \frac{2\mathbf{p}_{2(1)}}{4\pi\varepsilon_0 \varepsilon_d r_{12}^3}. \tag{32}$$

Following our prior work [73,74] we obtain the expression for the maximum fields near the nanoparticles

$$E_{\max,1(2)}^\omega = 2Q \frac{L(\omega)\beta + 2\beta^2 Q \left(\frac{a_{2(1)}}{r_{12}}\right)^3}{L^2(\omega) - 4\beta^2 Q^2 \left(\frac{a_1 a_2}{r_{12}^2}\right)^3} \overline{E}_\omega. \tag{33}$$

In the limit of $a_2 \approx 0$ $a_1 \approx r_{12}$, one gets

$$E_{\max,1}^\omega \approx \frac{2\beta Q}{L(\omega)} \overline{E}_\omega \, ; \; E_{\max,2}^\omega \approx \left[\frac{2\beta Q}{L(\omega)}\right]^2 \overline{E}_\omega. \tag{34}$$

As one can see in Fig. 4(a) the field is greatly enhanced in the vicinity of smaller particle. In our prior work [74], using more precise calculations we have shown that the simple analytical results (34) can be used as an upper bound on the field enhancement in the nanolens, or, as a matter of fact, in the nano-gap between two particles. In Fig. 4(b) we show the dimer of elliptical nanoparticles that resonates on our wavelength of choice of 1320 nm, as well as its extinction spectrum in Fig. 4(c). These results have been obtained using precise numerical calculations. So, the enhancement of the order of $Q^2$ for the asymmetric dimer is realistic and now we can see what it portends for the enhancement of nonlinearity.

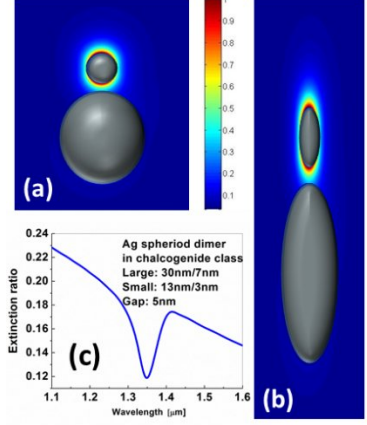

Fig. 4. (a) Spherical nanoparticle dimer with the electric field distribution. (b) Elliptical nanoparticle dimer resonant at 1320 nm and associate electric field distribution (c) Extinction spectrum of the above dimer

**4.b Effective nonlinearity of the plasmonic dimer**

The high field in the vicinity of the smaller nanoparticle will cause nonlinear polarization (9)

$$\boldsymbol{P}_{nl,2}^{\omega}(\boldsymbol{r},t) = P_{\max,2}^{\omega}\boldsymbol{G}_2(\boldsymbol{r})e^{-j\omega t} \tag{35}$$

where $\boldsymbol{G}_2(\boldsymbol{r})$ is the normalized distribution of nonlinear polarization near the smaller particle. Then, according (14) this polarization will induce the nonlinear dipoles of two particles

$$p_{nl,1}^{\omega} = 2\pi a_2^3 Q^2 \frac{2\beta\left(\dfrac{a_1}{r_{12}}\right)^3 \kappa P_{\max,2}^{\omega}}{L^2(\omega) - 4\beta^2 Q^2 \left(\dfrac{a_1 a_2}{r_{12}^2}\right)^3}$$

$$p_{nl,2}^{\omega} = 2\pi a_2^3 Q \frac{L(\omega)\kappa P_{\max,2}^{\omega}}{L^2(\omega) - 4\beta^2 Q^2 \left(\dfrac{a_1 a_2}{r_{12}^2}\right)^3}. \tag{36}$$

As one can see from comparison to (14) the nonlinear dipole of the larger nanoparticle 1 experiences additional enhancement relative to the dipole of the smaller nanoparticle 2. But note that now the volume of the smaller nanoparticle is present in the numerator of (36), hence the situation that is optimum for the external field enhancement in nanolens ,i.e. the limit of $a_2 \approx 0$  $a_1 \approx r_{12}$ is far from being optimal for the enhancement of nonlinear polarization.

Let us now estimate the effective nonlinear susceptibility of the nanolens. Finding from (19) and (33) the maximum nonlinear polarization near the smaller nanoparticle for the case of FWM and substituting it into (36) we obtain the nonlinear dipole of the larger particle 1 and then the effective third order susceptibility becomes

$$\chi_{eff}^{(3)} = 24 f \chi^{(3)} \kappa_3 \beta \ Q^5 \frac{\left[L(\omega)\left(\frac{a_2}{r_{12}}\right)^3 + 2\beta Q \left(\frac{a_1 a_2}{r_{12}^2}\right)^3\right]\left|L(\omega) + 2\beta Q \left(\frac{a_1}{r_{12}}\right)^3\right|^2}{\left[L^2(\omega) - 4\beta^2 Q^2 \left(\frac{a_1 a_2}{r_{12}^2}\right)^3\right]^2 \left|L^2(\omega) - 4\beta^2 Q^2 \left(\frac{a_1 a_2}{r_{12}^2}\right)^3\right|^2} \quad (37)$$

So, what is the maximum attainable nonlinearity enhancement? According to (34) the local field gets enhanced by a factor proportional to $Q^2$ instead of $Q$ for a single nanoparticle. For the Raman scattering, which is also a third-order nonlinear process, the enhancement with nanolens system can be $Q^8$ instead of $Q^4$ for a single nanoparticle, a tremendous improvement. But we cannot expect similar improvement for the FWM and other third order nonlinear processesbecause the largest enhancement of local fields is always attained when the volume smaller particle becomes negligibly small. But the key characteristic of (37), already noted above, is the presence of the volume of the smaller nanoparticle in the numerator, hence the optimum condition for maximum effective $\chi^{(3)}$ will not coincide with the condition maximum local field enhancement and overall enhancement will be much less than $Q^8$.

Optimizing (37) we find out that $\chi^{(3)}$ enhancement reaches its maximum when $4\beta^2 Q^2 \left(a_1 a_2 r_{12}^{-2}\right)^3 = 1/3$ and is equal to

$$\chi_{eff}^{(3)} \approx 5 f \chi^{(3)} \kappa_3 \beta^2 Q^6 . \quad (38)$$

Well, as one can see, the enhancement of $\chi^{(3)}$ and nonlinear index $n_2$ provided by the nanolens system is only proportional to the $Q^6$. This is rather easy to interpret. The local intensity in the nanolens gets enhanced by a factor proportional to $Q^4$, but then, the nonlinear refractive index change gets enhanced by the same additional factor $Q^2$, whether it is a single particle, nanolens, dimer, or nano-antenna. The additional enhancement provided by coupled particles composite (38) compared to the isolate nanoparticle composite (22) is about $\chi_{eff,2}^{(3)} / \chi_{eff,1}^{(3)} \approx 0.5 Q^2 / \beta$ i.e a factor on the order of 200. Overall enhancement for the previously considered case of $\beta \sim 1.35$ $Q \sim 20$ and $f = 0.001$ in chalcogenide glass can be as high as $3 \times 10^5$, but the relevant question is what does it mean in terms of maximum phase shift that can be obtain.

### 4.c Limitations of maximum phase shift with dimers

This shift can be obtained in a way similar to (26)

$$\Delta \Phi_{max} \approx 1.7 \kappa_3 \beta Q^5 \frac{n_2}{n_d} \overline{I}_0 . \quad (39)$$

Therefore the pump optical intensity required to achieve π-phase shift is $\overline{I}_\pi \sim 1.5 \times 10^7 \ W/cm^2$, i.e. only about 150 mW of peak power into 1μm$^2$ waveguide. This appears to be a reasonable power, but, of course the problem is that the local intensity is enhanced according to (33) roughly by $I_{max}/\overline{I}_\pi \sim 9\beta^4 Q^4 \approx 5 \times 10^6$, indicating that the local intensity will be on the scale of $I_{max} \approx 10^{14} \ W/cm^2$ which is way beyond the optical damage value. If we introduce once again the maximum local nonlinear index as $\Delta n_{max} = n_2 \overline{I}_{max} = 9\beta^4 Q^4 n_2 \overline{I}_0$, (39) can be re-written as

$$\Delta\Phi_{max} \approx 0.2\kappa_3 \frac{Q}{\beta^3} \frac{\Delta n_{max}}{n_d}. \tag{40}$$

This result for the nano-lens is even worse (by a factor of about 10) than the result (28) for the isolated nanoparticles. Clearly, the dependence $\kappa_4 Q$ is common to any type of nanostructure, monomer, dimer, trimer, or nano-antenna. The maximum achievable index of refraction $\Delta n_{max}$ changes the resonant SPP frequency which provides enhancement by the factor of $Q^2$ but then absorption coefficient also gets enhanced by the factor of $Q$ so only a single factor of $Q$ survives in the end. The factor in front of $\kappa_3 Q$ is reduced in dimers and more complicated structures relative to the monomers simply because a smaller fraction of the mode energy is contained in the region where the index change is maximal. Hence one should not expect any improvement in maximum obtainable nonlinear phase shift $\Delta\Phi_{max}$ beyond a single factor of Q in more complicated structures like trimers, bowtie antennae and so on, even if the effective nonlinear index can be enhanced beyond already huge enhancement in (38). Giant enhancement of nonlinearity will only mean that the nonlinear phase shift will saturate at much shorter distance but at essentially the same value of (28) or less, indicating that to the best of our knowledge with existing materials it is impossible to achieve true all-optical switching using plasmonic enhancement.

**5. Results and discussion**

In this section we illustrate the main results of our derivations and then make conclusions. Consider first Fig. 5(a) where the nonlinear phase shift $\Delta\Phi(z)$ induced by either XPM or SPM, is shown as a function of the propagation length in the chalcogenide glass waveguide (nonlinear index $n_2 = 10^{-13} \text{cm}^2/\text{W}$) doped with isolated Ag spheroids of Fig. 1(b). The incident pump power density is $\bar{I}(0) = 10^7 \text{W}/\text{cm}^2$ (corresponding to 100 mW of 1320 nm pump power into 1 μm² waveguide). The results are shown for four different filling density factors from f=10⁻⁶ to f=10⁻³ as well as for the pure chalcogenide waveguide without Ag spheroids (dashed line). As one can see, almost a three order of magnitude enhancement is achieved at short propagation length for densely filled waveguide with f=10⁻³ but the nonlinear phase shift saturates also at a very short distance of about 5μm with maximum phase shift of only 0.01rad. For lower filling factors initial enhancement is less but the saturation distance is also longer due to reduced absorption and in the end the maximum phase shift saturates at the same value of 0.01 rad. For the un-enhanced waveguide the saturation does not occur (the absorption length in the waveguide is longer than 1cm) and as one can see for long propagation distances the un-enhanced structure outperforms all the plasmoncially enhanced ones.

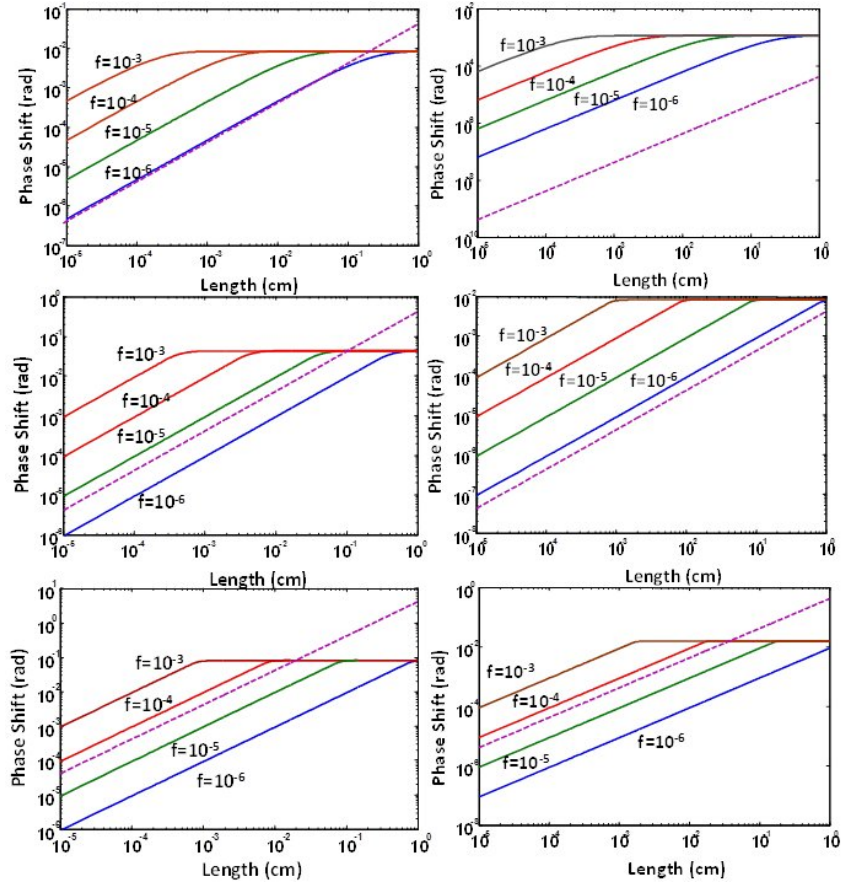

Fig. 5. Nonlinear phase shift in the chalcogenide waveguide doped with isolated Ag spheroids (a-c) and the dimers (d-f) with different filling factors f at various intensities. The phase shift of undoped waveguides is shown by dashed line.

If we increase the input power by a factor of 10 to $\overline{I}(0) = 10^8 \, \text{W/cm}^2$ the local power density near the metal surface will reach roughly $3 \times 10^{11} \text{W/cm}^2$ which should lead to local index change of 3%. This index change is clearly unattainable. First of all, optical damage will most probably ensue, but even in absence of it, the index change will saturate at a value that is less than 1% [72]. So, to be optimistic, we disregard the possibility of optical damage but still consider the saturation of nonlinear change with the results seen in Fig. 5(b). Once again, significant enhancement can be obtained at small propagation distances, but the at saturation the maximum phase shift is still less than 0.05 rad

When the input power is increased by another factor of 10 to $\overline{I}(0) = 10^9 \, \text{W/cm}^2$ (more than 10W of peak power) as shown in Fig. 5(c) the saturation still prevents nonlinear index change from exceeding 0.1rad although this change takes place over propagation distance of no more than 10 μm. Of course, we need to stress here that in real structures optical damage most probably will occur at the local power densities in excess of TW/cm² that would occur near the metal surface. Note that for the waveguides without nanoparticles nonlinear phase shift of π radians required for switching is achieved at a few mm propagation distance.

We now turn our attention to the chalcogenide waveguides doped with optimized Ag dimers and first consider them at relatively low input power density of just $\overline{I}(0) = 10^4 \, \text{W/cm}^2$ with the results shown in Fig. 5(d). As one can see at small propagation distances nonlinearity gets enhanced by more than 5 orders of magnitude and appreciable phase change of 0.001 rad is achieved at propagation distance of only 10 μm, but this is where it gets saturated.

Further increase of input power density to $\overline{I}(0) = 10^6 \text{ W/cm}^2$ (Fig. 5e) and $\overline{I}(0) = 10^8 \text{ W/cm}^2$ [Fig. 5(f)] does not lead to significant increase in the maximum nonlinear phase shift – it remains below 0.02rad and thus insufficient for optical switching.

Moving on to the frequency conversion by means of FWM we first consider the same waveguide doped with isolated Ag spheroids and plot the conversion efficiency vs. distance in Fig. 6(a) for input pump power of $\overline{I}(0)=10^7 \text{W/cm}^2$. Once gain we obtain tremendous enhancement of conversion efficiency at short distances. In only a few micrometers one can attain conversion efficiency of nearly 0.01% (-40dB) which can be sufficient for some applications, but probably not for frequency conversion in optical communication schemes or for optical switching. At longer distances the conversion efficiency deteriorates due to absorption. Note that one can adjust the distance at which maximum conversion efficiency is achieved by varying f.

Going to the waveguides doped with dimers [Fig. 6(b)] allows one to reach conversion efficiency enhancement of nearly 10 orders of magnitude for very short structures, but the absolute value of conversion efficiency is not bound to exceed 0.0001%. Perhaps this conversion efficiency is sufficient for some specialized operation, such as generating of entangled pairs of photons or autocorrelation measurements, but it is not enough for signal processing.

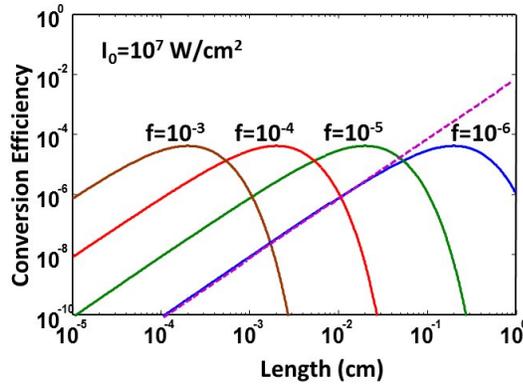

Fig. 6. Conversion efficiency in the FWM in the waveguide doped with isolated Ag spheroids (a) and the dimers (d-f) with different filling factors f. The phase shift of undoped waveguides is shown by dashed line.

## 6. Conclusion

Our conclusions are two-fold. On the one hand, using waveguides impregnated with metallic monomers, dimers, and other constructions (one may call them plasmonic metamaterials) allows one to achieve huge enhancement of effective nonlinear index, up to the order of $10^5$ and more due to high degree of field concentration in the "hot spots". On the other hand, strong absorption in the metal causes saturation of the nonlinear phase shift for SPM and XPM or frequency conversion efficiency in case of FWM and OPG at very short distances. Given the fact that maximum local index change is limited, generously, to about 1% due to optical damage, the nonlinear phase shift saturates at a very small value of a few tens of milliradians – which is insufficient for any photonic switching operation. Similarly, conversion efficiency saturates at values of less than -40dB making use of plasmonic nonlinear metamaterials for this purpose highly inefficient. It is also clear that changes in Q by a factor of 2 or 2 -that might be attainable in silver (although not demonstrated to date) due to improvement in fabrication methods, will not change these conclusions in any substantial way, and only assure earlier saturation of the nonlinear conversion. The one and only advantage of nonlinear plasmonic metamaterial is that nonlinear effects may be observable at very small propagation distances of a few micrometers with reasonable (but not low!) optical powers. Yet there is a big difference between being observable and being practical and at this point, with available metals and nonlinear materials, one cannot see how the nonlinear plasmonic metamaterials can bridge this gap.

This conclusion is in line with our general conclusions about utility of plasmonic enhancement – the devices that have inherently low efficiency (e.g. Raman sensors ) can be enhanced spectacularly, with important implications for sensing, but the devices that are already reasonably efficient (LED, Solar cell, etc) will only see their performance deteriorate when metal is introduced. The nonlinear devices are no different – when propagation distance is short, very low efficiency can be enhanced significantly, but it will remain disappointingly low. For the longer devices the performance will deteriorate.

In fact, these conclusions about prospects of using plasmonic resonances to enhance nonlinearity do not appear to be surprising at all. Various resonant schemes for enhancement nonlinearity have been studied at length. Some of the schemes rely upon intrinsic material resonances; others try to take advantage of photonic resonant structures, such as micro-resonators and photonic crystals. The Q-factor of the resonances ranges from a few hundreds to tens of thousands, and yet in the end, none of the resonant schemes had found practical applications to this day, due to the fact that resonance is always associated with excessive absorption and dispersion. To this day optical fiber remains the nonlinear medium of choice in which low nonlinear coefficients are more than compensated by the long propagation length and high degree of confinement.. The only other media in which all-optical switching has been consistently demonstrated is semiconductor optical amplifier (SOA) in which the loss simply does not exist due to optical gain. Neither fiber nor SOA relies upon any resonance despite its apparent appeal - one always has a lot to gain by avoid absorption and excessive dispersion associated with resonance.

So, if the numerous relatively high $Q$ resonant schemes for enhancing optical nonlinearity have not become practical, it would be naive to expect plasmonic resonance in metal nanoparticles with $Q$ barely of the order of 10 to succeed with higher Q schemes have failed. Thus in retrospect one can say this work only confirms the obvious. And yet this obvious fact has not been universally accepted by the community, and it is our hope that our effort has been useful as it has revealed the nature and limitations of the plasmonic enhancement of $\chi^{(3)}$ in great detail and without reliance on excessive numerical modeling.

Not to end on entirely pessimistic note, we should mention that there exist broad classes of nonlinearities that rely on temperature change and which the index change in excess of 1% can be attained. That includes both conventional thermo optical effects in standard materials such as Si [75] and the thermally induced metal-to-insulator transition in materials like $VO_2$ [76]. The index change in the latter is on the order of 1! But the switching time is determined by the heat transfer and is typically very slow. It is here where in our opinion plasmonic can shift the whole paradigm since if the local heating can be reduced to a nanometer scale (which is of course the case for nm scale particles) then the heat diffusion time would be on the order of picoseconds and one could talk about *ultrafast thermal nonlinearities!* We shall explore this idea as well as using non-metallic structures with negative permittivity and lower loss in the future publications.

To conclude it was not our intent to make any predictions of where this research will go in the future, the sole purpose of this work, was to provide a set of simple expressions and numbers for the others so they can ascertain the prospects for using nonlinear plasmonic metamaterials for their own applications. Still we may make a broad statement, that plasmonically enhanced structures in nonlinear optics might not find too many applications requiring decent efficiency, such as switching, wavelength conversion, etc, but may be of great use in such applications where efficiency is not much of an issue such as sensing and also fundamental studies of optical properties of different materials under extremely high fields.

**Acknowledgments**
This work was supported by the Air Force Office of Scientific Research under the contract FA9550-10-1-0417 and Mid-InfraRed Technologies for Health and the Environment (MIRTHE) Research Center (National Science Foundation—Engineering Research Centers)